\documentclass[a4paper]{jpconf}
\usepackage{graphicx}

\newcommand{\ba}{\begin{eqnarray}}
\newcommand{\ea}{\end{eqnarray}}
\newcommand{\beqs}{\begin{eqnarray}}
\newcommand{\eeqs}{\end{eqnarray}}

\begin{document}
\title{Electromagnetic and gravitomagnetic structure of pions and pion-nucleon scattering } 

\author{O.V. Selyugin}

\address{ BLTPh, JINR,  Dubna, Russia}

\ead{selugin@theor.jinr.ru}

\begin{abstract}
Taking into account  PDFs obtained by various Collaborations,
 the momentum transfer dependence of GPDs of the pion
 are obtained. The calculated   electromagnetic and gravitomagnetic
  form factors of the  pions and nucleons are used for the description
  of the  pion-nucleon   elastic scattering in a wide energy and momentum transfer region
  with minimum fitting parameters.
\end{abstract}

\section{Introduction}

%
%
%
%
%

     The remarkable property of $GPDs$ is that the integration of  different momenta of
     GPDs over $x$ gives us  different hadron form factors  \cite{Mil94}.  
     The $x$ dependence of $GPDs$ is in most part  determined by the standard
    PDFs, which are obtained by  Collaborations from the analysis of
    the deep-inelastic processes.
%


Let us modify the original Gaussian ansatz
 and choose  the $t$-dependence of  GPDs  in a simple form
$ {\cal{H}}^{q} (x,t) \  = q(x) \   exp [  a_{+}  \
   f(x) \ t ],                                     $
  with $f(x)= (1-x)^{2}/x^{\beta}$
   \cite{ST-PRDGPD}.
The isotopic invariance can be used to relate the proton and neutron GPDs.


   The complex analysis of the corresponding description of the electromagnetic form factors of the proton and neutron
    by  different  PDF sets  (24 cases) was carried out in \cite{GPD-PRD14}. These
   PDFs include the  leading order (LO), next leading order (NLO) and next-next leading order (NNLO)
   determination of the parton distribution functions. They used the different forms of the $x$ dependence of  PDFs. 
    We slightly complicated the form of GPDs  in comparison with the equation used in     \cite{ST-PRDGPD},  
   but it is the simplest one as compared to other works (for example \cite{DK-13}).
%
%
%
%

 The hadron form factors will be obtained  by integration  over $x$ in the whole range $0 - 1$.
 Hence, the obtained  form  factors will be dependent on the $x$-dependence of the forms of PDF
 at the ends of the integration region.
    On the basis of our GPDs with PDFs
    ABM12 \cite{ABM12}    we calculated the hadron form factors
     by the numerical integration
   and then
    by fitting these integral results by the standard dipole form with some additional parameters
$   F_{1}(t)  = (4m_p - \mu t)/(4m_p -  t ) \  \tilde{G}_{d}(t) $
  with $ \tilde{G}_{d}(t) = 1/(1 + q/a_{1}+q^{2}/a_{2}^2 +  q^3/a_{3}^3)^2 $ that is slightly  different from
  the standard dipole form on two additional terms with small sizes of coefficients.
  The matter form factor 
\ba
 A(t)=  \int^{1}_{0} x \ dx
 [ q_{u}(x)e^{2 \alpha_{H} f(x)_{u} / t  } 
  + q_{d}(x)e^{ 2 \alpha_{H} f_{d}(x)  / t}  ] 
\ea
 is fitted   by the simple dipole form  $  A(t)  =  \Lambda^4/(\Lambda^2 -t)^2 $.
        These form factors will be used in our model of the proton-proton and proton-antiproton elastic scattering
        and  further in one of the vertices of the pion-nucleon elastic scattering.

%
%
%

%
%
%

\section{Hadron form factors and elastic nucleon-nucleon scattering}

        Both hadron  electromagnetic and gravitomagnetic form factors were used
        in the framework of the high energy generalized structure   (HEGS)
        model  of the elastic nucleon-nucleon scattering.
        This allows us to build a model with a minimum number of fitting
        parameters \cite{HEGS0,HEGS1,NP-hP}.

  The model is very simple from the viewpoint of the number of fitting parameters and functions.
  There are no any artificial functions or any cuts which bound the separate
  parts of the amplitude by some region of momentum transfer.
        In the framework of the model the description of the experimental data was obtained simultaniously
        at the large momentum transfer and in the Coulomb-hadron region, using CNI phase \cite{selmp1,Selphase},
        in the energy region from $\sqrt{s}=9 $ GeV
        up to LHC energies.  The model gives a very good quantitative description of the latest
        experimental data at $\sqrt{s}=13$ TeV \cite{Sel-PL19}.

\section{GPDs of pion }

      The pion structure in some sense is simpler than the nucleon structure.
 In the nucleon there are 3 constituent quarks that can create  different configurations, for example,
  such as "Mercedes star" or a linear structure
  with a quark on one end and a di-quark on the other.
  These configurations can lead to   different results for hadron interactions,
  for example, the Odderon-hadron coupling.
  For a meson we have only two quark states
$$ |\pi> =|q\bar{q}> + |q\bar{q} \ q\bar{q}> + |q\bar{q} \ g> ..... $$
  It is needed to note that the standard definition of the pion form factor
  through the matrix elements of the electromagnetic vector current
  gives
\ba
 \langle   \pi^{+}(\vec{p^{'}}) |V_{\mu}(0)| \pi^{+}(\vec{p})\rangle
  = (p_{\mu}^{'} + p_{\mu}) F_{\pi}(Q^{2}),
\ea
  with $Q^{2}=-q^2$ and $F_{\pi}(Q^{2})$ being the space-like form factor of a pion  \cite{ETM}.

%

 We have focused on the zero-skewness limit, where  GPDs
have a probability-density interpretation in the longitudinal Bjorken x and
the transverse impact-parameter distributions,
 \ba
{\cal{H}}^{q} (x,t) \  = q(x)  \   e^{2 a_{H}   f(x)_{q}  \ t };  \ \ \ 
{\cal{E}}^{q} (x,t) \  = q(x) (1-x)^{\gamma} \   e^{2 a_{E}  \  f(x)_{q}  \ t }; \ \ \  
\label{t-GPDs-E}
\ea
 where
 $ f_{q}(x) =  \frac{(1-x)^{2+\epsilon_{u}}}{(x_{0}+x)^{m}}$.

 Hence, the obtained  form  factors will be dependent on the $x$-dependence of the forms of PDF at the ends of the integration region.
  Some  PDFs  have the polynomial form of $x$ with
     different power.  Some other have the exponential dependence of $x$.
  As a result, the behavior of  PDFs, when $x \rightarrow 0$ or $x \rightarrow 1$,  can  impact  the
    form of the calculated form factors.

 Different Collaborations have determined the  PDF sets  from the inelastic processes only in  some region of $x$, which is only
 approximated to $x=0$ and $x=1$.
  Also, there is a serious problem in determining  the main
  ingredient of GPDs of a pion - the form of the parton distribution functions.
   The predictions based on the perturbative QCD and the calculations using the Dyson-Schwinger equation
   lead to  $( 1-x)^2$ at $x \rightarrow 1 $.
   However, the constituent quark model and calculation in the framework of
   the Nambu-Jona-Lasino model lead to the linear behavior $(1-x)_{x \rightarrow 1}$.
   Several next-to-leading order (NLO) analyses of the Drell-Yan data show
   that the valence distribution turned out to be rather hard at high momentum fraction x ,
typically showing  only a linear or slightly faster falloff.
  Correspondingly, there are many different forms of the PDF of a pion. 

   We examine many of them and keep two PDFs leading to approximately
  the same results: one is (Mezrag et al. (2016) \cite{Mez16} )    and other is   \cite{RayT}

\begin{figure}
\begin{center}
\includegraphics[width=.49\textwidth]{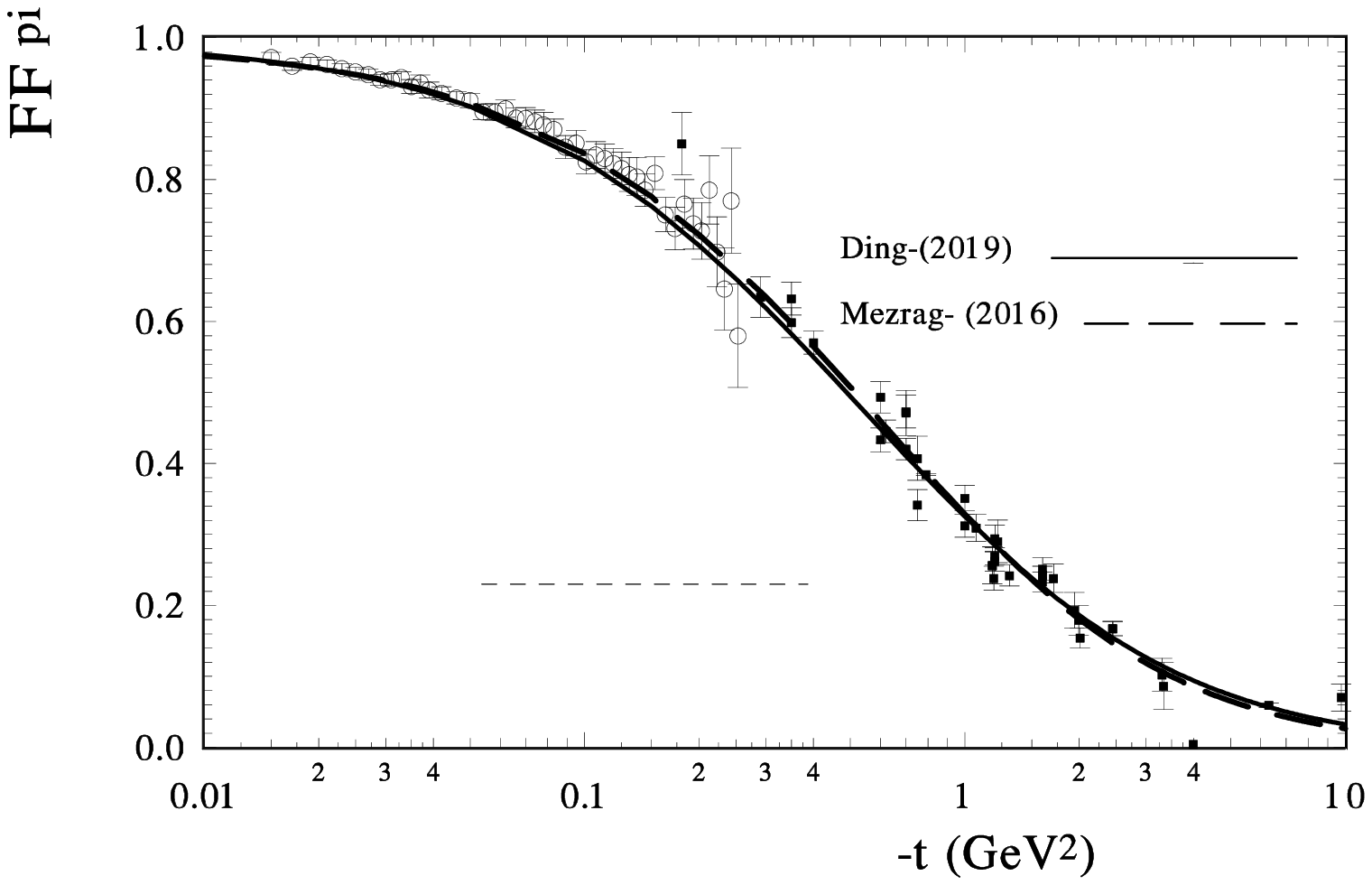} 
\includegraphics[width=.49\textwidth]{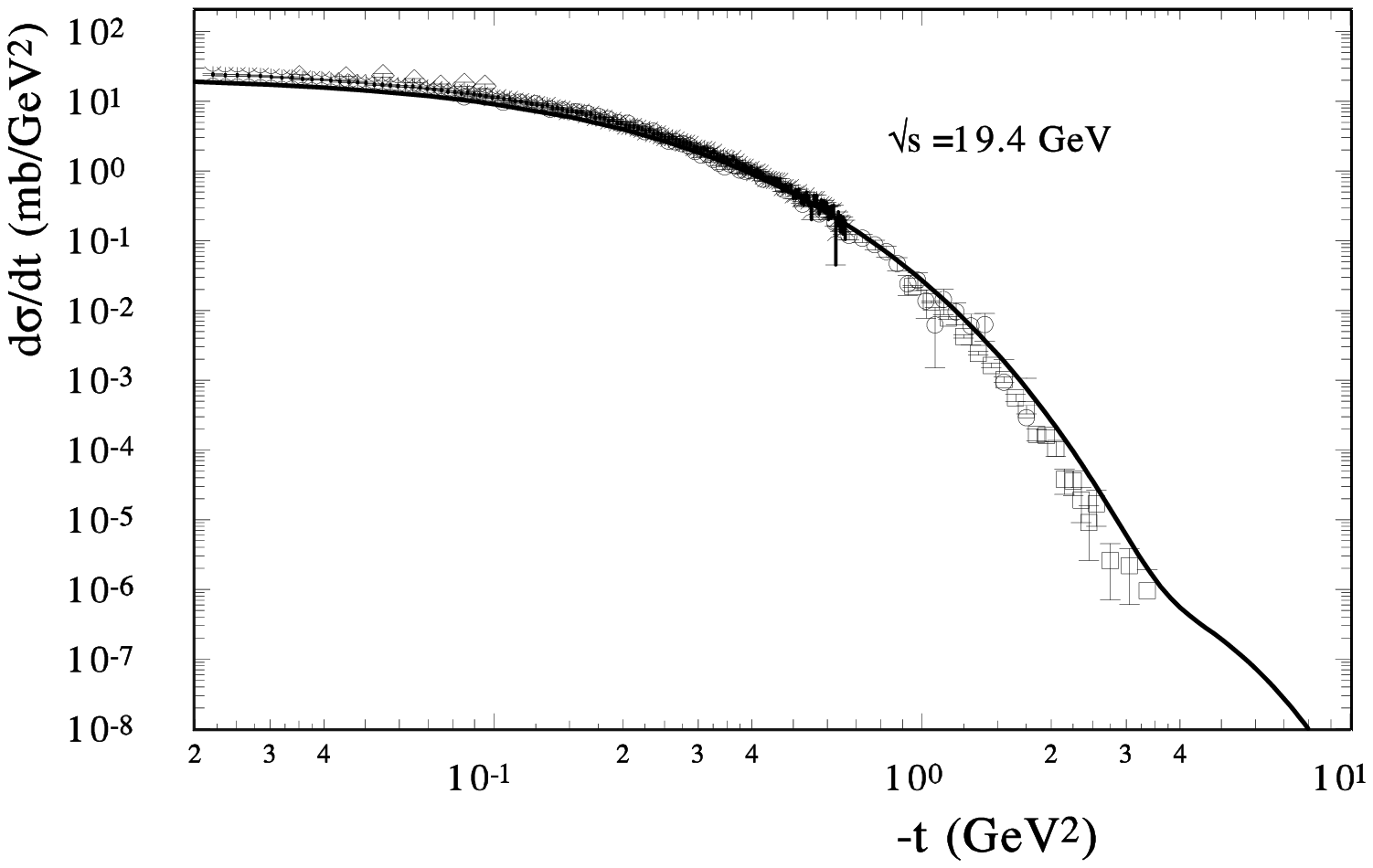} 
\vspace{1.cm}
\caption{a) [left] The electromagnetic form factor of the $\pi$-meson
(hard and dashed curves our calculations with  PDF \cite{Mez16} and  and \cite{RayT}, respectively;
 the circles and squares - the experimental data \cite{ff1} \\
 b) [right] The differential cross sections of the elastic scattering of $\pi^{+}p$
 at $\sqrt{s} = 19.4 $ GeV (curves - our model calculations, circles, quires, triangles up and triangles down - \cite{Rub19am}
}
\end{center}
\label{Fig_}
\end{figure}

 On the basis of our GPDs with
    PDFs    we have calculated the pion form factors
     by the numerical integration
   and then
    by fitting these integral results by the standard monopole form
  and obtained $\Lambda^{2}_{\pi} =0.55$.
  In Fig.1a, the comparison of our calculation with the existing experimental data of the
  pion form factor are presented. It is seen that the difference between the calculations
  of  our two chosen  PDFs is small. 

   The matter form factor $A_{Gr}^{\pi}(t)$ is calculated as the second Mellin momentum
\ba
 A_{Gr}^{\pi}(t)=  \int^{1}_{0} x \ dx \
 q_{\pi}(x)e^{2 \alpha_{\pi} f(x) / t  } 
\ea
 and is fitted   by the simple dipole form  $  A(t)  =  \Lambda^4/(\Lambda^2 -t)^2 $.
        These form factors will be used in our model of the $\pi^{+}p$ and $\pi^{}p$  elastic scattering.
 Our calculations of the second momentum of GPDs of a pion
  are shown  
  that the impact of different PDFs is tangible only
  at large momentum transfer.

 \section{Hadron form factors and the elastic pion-nucleon scattering }

  Let us determine the Born terms of the elastic pion-nucleon scattering amplitude
  in the same form as we determined the elastic nucleon-nucleon scattering
   amplitudes. Using  both the (electromagnetic and gravitomagnetic) form factors
   of a pion and a nucleon, we obtain  two asymptotic terms:
  \begin{eqnarray}
 F_{mh}^{Born}(s,t)=&& h_1 \ G_{N}(t)  \ G_{\pi}(t) \ F_{a}(s,t)    +  h_{2} \  A_{N}(t) \ A_{\pi}(t) \  F_{b}(s,t)
    \label{FB}
\end{eqnarray}
  where $F_{a}(s,t)$ and $F_{b}(s,t)$  have the standard Regge form: 
$$ F_{a}(s,t) \ = (\hat{s}^{\epsilon_1} \ + \ r_{a}/\sqrt{\hat{s}}) \ e^{B(\hat{s}) \ t}; \ \ \
 F_{b}(s,t) \ = (\hat{s}^{\epsilon_1} + \ r_{b}/\sqrt{\hat{s}} )  \ e^{B(\hat{s})/4 \ t}, $$
 $   \hat{s}=s \ e^{-i \pi/2}/s_{0}$ ;  $s_{0}= 1 \ {\rm GeV^2}$, and
   intercept $1+\Delta_{1} =1.11$ was chosen the same as for the nucleon-nucleon elastic
  scattering. Hence, at the asymptotic energy we have the universality
  of the energy behavior of the elastic hadron scattering amplitudes. 
 The slope of the scattering amplitude having the standard logarithmic dependence on the energy
 $   B(s) = \alpha^{\prime} \ ln(\hat{s}) $
  with $\alpha^{\prime}=0.24$ GeV$^{-2}$ is again the same size as for the nucleon-nucleon elastic scattering
  Examining the pion-nucleon elastic scattering at low energies, we take into
  account the contributions of the non-leading Reggions using the  form factors of
  the pion and nucleon.  
%
%
%
The final elastic  hadron scattering amplitude is obtained after unitarization of the  Born term.
    So, at first, we have to calculate the eikonal phase
 $ \chi(s,b) \   =  -\frac{1}{2 \pi}
   \ \int \ d^2 q \ e^{i \vec{b} \cdot \vec{q} } \  F^{\rm Born}_{h}(s,q^2)  $ and then obtain the final hadron scattering amplitude. 


 We take into account the experimental data on the  $\pi^{+} p$ and $\pi^{-} p$
    elastic scattering from $\sqrt{s} = 7.807$  GeV up to the maximum measured
    at $\sqrt{s} = 25.46$ GeV. The total number of the experimental data $N_{exp.}=2022$.
    As in the case of the nucleon scattering, we take into account in
    the fitting procedure the statistical and systematic errors separately.
    Only the statistical errors are included in the standard calculations of
    $\chi^2$. The systematic errors are taken into account as some additional normalization
    of the experimental data of the separate set. As a result, we have obtained
    $\chi^2/n_{d.o.f.} = 1.2$. In Fig. 1b the model calculations are
    compared with the elastic  $\pi^{+} p$ .

\section{Conclusion}

 We have examine the new form of the momentum transfer dependence of GPDs of hadrons to obtain
       different form factors, including
     Compton form factors, electromagnetic form factors,
     transition form  factor and gravitomagnetic form factor.
Our model of GPDs,
  based on the analysis of  practically all existing experimental data on the
    electromagnetic form factors of the proton and neutron,
 leads to a good description
of the proton and neutron  electromagnetic form factors  simultaneously.
The chosen form of the momentum transfer dependence of GPDs of the pion
(the same as t-dependence of nucleon) allows us to describe the electromagnetic form factor of pions
and obtain the pion gravitomagnetic form factors.
As a result, the description of different reactions based on
     the same   representation of the hadron structure was obtained.
The model opens up a new way to determining the true form of the GPDs and      hadrons structure.



\end{document}